\documentstyle[12pt,epsf]{article}

\begin{document}
\begin{flushright}
\mbox{BA-99-49}\\
\mbox{FERMILAB-Pub-99/159-T}\\
\mbox{hep--ph/9906297}\\
\mbox{June 1999} \\[0.4in]
\end{flushright}
\begin{center}
	{\large \bf Bimaximal Mixing in an $SO(10)$ Minimal Higgs Model
	\\[0.5in]}
Carl H. Albright$^1$ \\ Department of Physics \\ Northern
Illinois University, DeKalb, IL 60115 \\
and Fermi National Accelerator Laboratory \\
P.O. Box 500, Batavia, IL 60510 \\[0.2in]
S.M. Barr$^2$ \\
Bartol Research Institute \\ University of Delaware \\
Newark, DE 19716\\[0.3in]
\end{center}

\begin{abstract}

An $SO(10)$ SUSY GUT model was previously presented based on a minimal 
set of Higgs fields.  The quark and lepton mass matrices derived fitted the
data extremely well and led to large $\nu_{\mu} - \nu_{\tau}$
mixing in agreement with the atmospheric neutrino data and to the small-angle 
MSW solution for the solar neutrinos.  Here we show how a slight
modification leading to a non-zero up quark mass can result in bimaximal mixing
for the atmospheric and solar neutrinos.  The ``just-so'' vacuum solution
is slightly favored over the large-angle MSW solution on the basis of the 
hierarchy required for the right-handed Majorana matrix and the more
nearly-maximal mixing angles obtained.\\[0.2in]

\end{abstract}
$\line(75,0){75}$\\
\noindent
$^1$E-mail: albright@fnal.gov; $^2$E-mail: smbarr@bartol.udel.edu\\
\thispagestyle{empty}
\newpage

Atmospheric neutrino data \cite{atm} reveal a large and, in fact, nearly maximal
mixing between $\nu_{\mu}$ and some other neutrino, which most
plausibly is assumed to be $\nu_{\tau}$, though it may also be a
sterile neutrino. The solar neutrino problem \cite{solar}, on the other hand, 
can be solved either by a small mixing 
($\sin^2 2\theta_{e \mu} \sim 6 \times 10^{-3}$) or by a nearly
maximal mixing of $\nu_e$ with $\nu_{\mu}$.  
In the former case one has the small-angle MSW 
solution \cite{MSW}, while in the latter case one has either the large-angle 
MSW solution \cite{MSW} or the vacuum oscillation solution \cite{vac}. Cases 
where both the atmospheric and solar neutrino anomalies are solved by nearly
maximal mixing \cite{bimax} are often called ``bimaximal.'' 

There have been several interesting suggestions about how bimaximal
mixing might arise theoretically \cite{bimaximal}. Here we suggest a scheme 
that has certain novel features. The basic idea, which at first sounds 
artificial but which actually can emerge quite simply 
and naturally as will be seen, is that the large mixing of $\nu_{\mu}$
and $\nu_{\tau}$ originates from transformation of the {\it charged lepton} 
mass matrix,
whereas the large mixing of $\nu_e$ and $\nu_{\mu}$ originates in the
{\it neutrino} mass matrix itself. (Of course, it only makes sense to draw
this distinction if there is a preferred basis of families, which
is here provided by some underlying theory of flavor.)
The idea that the large $\nu_{\mu}-\nu_{\tau}$ mixing arises from the 
mass matrix of the charged leptons was proposed in \cite{ab1}, where
it emerged as part of a complete model for the quark and lepton masses and
mixings.
One of the virtues of this idea is that it allows a simple resolution 
of the supposed paradox that the mixing of the second and third families
is small for the quarks, $V_{cb} \cong 0.04$, and large for the leptons,
$V_{\mu 3} \simeq 0.7$.

The model developed in \cite{ab1} was only a model for the heaviest two 
families with 
the first family being approximated as massless. In \cite{ab2} this model
was extended to the first family in a very economical way that gave
several additional predictions, among which were that the 
$\nu_e-\nu_{\mu}$ mixing angle is small, and in fact precisely in the 
presently allowed range for the small-angle MSW solution. This small
$\nu_e-\nu_{\mu}$ angle, like the large $\nu_{\mu}-\nu_{\tau}$ angle,
arose from diagonalization of the {\it charged} lepton mass matrix.

In the present paper we show that a slight alteration of that 
model leads to an equally predictive scheme that has bimaximal mixing.
All the predictions for quark and charged lepton masses, for the CKM parameters,
and for the $\nu_{\mu}-\nu_{\tau}$ mixing are left essentially unaffected;
however, the $\nu_e-\nu_{\mu}$ mixing moves from the small-angle MSW
value to become maximal. We will first very briefly review the model of 
\cite{ab1} and \cite{ab2}, noting the features relevant to lepton mixing, 
and then proceed to show how bimaximal mixing can arise in it.

The model is based on supersymmetric $SO(10)$ and leads to the
following matrices: $U^0$ for up-type quarks, $D^0$ for down-type quarks,
$L^0$ for the charged leptons, and $N^0$ for the Dirac neutrino masses, 
where the superscript 0 refers to the matrices at the unification
scale.

\begin{equation}
\begin{array}{ll}
U^0 = \left( \begin{array}{ccc}
0 & 0 & 0 \\
0 & 0 & \epsilon/3 \\
0 & - \epsilon/3 & 1 \end{array}
\right) M_U, & D^0 = 
\left( \begin{array}{ccc}
0 & \delta & \delta' \\
\delta & 0 & \sigma + \epsilon/3 \\
\delta' & - \epsilon/3 & 1 \end{array}
\right) M_D, \\
& \\
N^0 = \left( \begin{array}{ccc}
0 & 0 & 0 \\
0 & 0 & - \epsilon \\
0 &  \epsilon & 1 \end{array}
\right) M_U, & L^0 = 
\left( \begin{array}{ccc}
0 & \delta & \delta' \\
\delta & 0 &  - \epsilon \\
\delta' & \sigma + \epsilon & 1 \end{array}
\right) M_D. 
\end{array}
\end{equation}

\noindent
These matrices arise from simple diagrams in the $SO(10)$ unified model, which 
involve only five effective Yukawa terms that have the
forms $({\bf 16}_3 {\bf 16}_3) {\bf 10}_H$,  $({\bf 16}_2 {\bf 16}_3)
{\bf 10}_H {\bf 45}_H$,  $[{\bf 16}_2 {\bf 16}_H]$ $[{\bf 16}_3
{\bf 16}'_H]$,  $[{\bf 16}_1 {\bf 16}_2] [{\bf 16}_H
{\bf 16}'_H]$, and $[{\bf 16}_1 {\bf 16}_3] [{\bf 16}_H
{\bf 16}'_H]$. These lead, respectively, to the entries in Eq. (1) 
that are denoted $1$, $\epsilon$, $\sigma$, $\delta$, and $\delta'$.
The numerical subscripts are family indices, while the subscript
H denotes a Higgs multiplet. The notation $[{\bf 16} \; {\bf 16}]$ implies
the spinors are contracted into an $SO(10)$ vector. The VEV, $\langle
{\bf 16}_H \rangle \sim M_G$, lies in the $SU(5)$ singlet direction and
helps break $SO(10)$ down to the Standard Model,
while $\langle {\bf 16}'_H \rangle \sim M_W$ lies in the weak doublet
direction and helps break the electroweak interactions. The expectation value
of the adjoint ($\langle {\bf 45}_H \rangle$) is proportional to the $SO(10)$ 
generator $B-L$, as required for the Dimopoulos-Wilczek mechanism to provide 
the doublet-triplet splitting \cite{2-3}. The foregoing information is
sufficient to derive the matrices in Eq. (1).

The family hierarchy results from the smallness of the parameters
$\epsilon \cong 0.14,\ \delta \cong 0.008$, and $\left|
\delta' \right| \cong 0.008$. The parameter $\sigma \cong 1.8$
is not small, however, and is the key to understanding many of the 
qualitative and quantitative features of the quark and lepton spectrum,
including the fact that the second and third families of leptons
have a large mixing while that of the quarks is small.  The point is 
that $SU(5)$ relates left-handed (right-handed) down quarks to 
right-handed (left-handed) charged leptons, and consequently relates the
$ij$ element of $D^0$ to the $ji$ element of $L^0$.  That is why in Eq. (1)
the large parameter $\sigma$ appears in $L^0_{32}$ where it leads to large
$\mu^- - \tau^-$ mixing and hence large $\sin^2 2\theta_{\mu\tau}$, whereas
it does not appear in $D^0_{32}$ where it would give large $V_{cb}$,
but rather appears in $D^0_{23}$ where it affects only the unobservable
mixing of right-handed quarks.  The great difference in
magnitude between $V_{cb}$ and $\sin^2 2\theta_{\mu\tau}$ is thus a 
consequence of $D^0$ and $L^0$ being highly asymmetric and the peculiarities
of $SU(5)$ invariance.  

As shown in \cite{ab1} and \cite{ab2}, the matrices in Eq. (1) give a 
remarkably good fit to all the known quark and lepton masses and mixings, 
in particular, when the small-angle MSW solution \cite{MSW} is relevant for 
the solar neutrino problem.  We now review the lepton results and 
explore the possibility of large-angle $\nu_e - \nu_{\mu}$ mixing in this 
$SO(10)$ unified model. 

The lepton mixing matrix is given by

\begin{equation}
V_{lepton} = U_L^{\dag} U_{\nu},
\end{equation}

\noindent
where $U_L$ is the unitary transformation of the left-handed charged leptons 
required to diagonalize $L^0$, and $U_{\nu}$ is the complex orthogonal
transformation of the left-handed neutrinos required to diagonalize the 
light-neutrino mass matrix, $M_{\nu} = - N^T M_R^{-1} N$.  In $SO(10)$ 
the Dirac neutrino mass matrix $N^0$ and up quark mass matrix $U^0$ are related,
and as given in Eq. (1) have vanishing first rows and first columns. 
This is to be regarded as only an approximation to the real world, but
as we shall see later, it is a very good approximation.

With this particular texture for $N^0$, no matter what form $M_R$ assumes, the 
light-neutrino mass matrix $M_{\nu}$ will also have vanishing first row and 
column, and we can write

\begin{equation}
U_{\nu} = \left( \begin{array}{ccc}
1 & 0 & 0 \\
0 & c & s \\
0 & -s & c 
\end{array}
\right),
\end{equation}

\noindent
where the parameters $s$ and $c$ are complex, in general, with
$c^2 + s^2 = 1$. From the form of $N^0$ it is easy to see that,
formally speaking, $\left| s \right| = O(\epsilon)$. If the unknown 
matrix $M_R$ is parametrized by $(M_R^{-1})_{ij} = a_{ij} \Lambda^{-1}_R$,
then, in units of $(M_U^2/\Lambda_R)$, one has $(M_{\nu})_{1j} = 
(M_{\nu})_{j1} = 0$, $(M_{\nu})_{22} = \epsilon^2 a_{33}$, 
$(M_{\nu})_{23} = (M_{\nu})_{32} = \epsilon a_{33} - \epsilon^2 a_{23}$, 
and $(M_{\nu})_{33} = a_{33} - 2 \epsilon a_{23} + \epsilon^2 a_{22}$.
This gives 

\begin{equation}
\tan 2 \theta^{\nu}_{23} \equiv 2sc/(c^2 - s^2) =
2 \epsilon \left( \frac{a_{33} - \epsilon a_{23}}{a_{33} - 2 \epsilon
a_{23} + \epsilon^2 a_{22} - \epsilon^2 a_{33}}\right),
\end{equation}

\noindent
and thus $\theta^{\nu}_{23} \sim \epsilon$, unless the parameters 
$a_{ij}$ are fine tuned to have a special relationship to each other.

We see then, that the contributions to the leptonic mixings
coming from $U_{\nu}$ are either zero or small. The leptonic mixings thus 
arise from $U_L$. This is good news, since $L^0$ in Eq. (1) is
known in this model.  With the values of the
parameters $\epsilon$, $\sigma$, $\delta$, and $\delta'$ given earlier 
as determined by fitting known quantities, one finds that

\begin{equation}
U_L^{\dag} = \left( \begin{array}{ccc}
c_{12} & -s_{12} c_{23} & s_{12} s_{23} \\
s_{12} & c_{12} c_{23} & -c_{12} s_{23} \\
0 & s_{23} & c_{23} 
\end{array} \right),
\end{equation}

\noindent
where

\begin{equation}
\begin{array}{l}
\tan \theta^L_{23} = s_{23}/c_{23} \cong \sigma + \epsilon \cong 1.9, \\
\\
\sin \theta^L_{12} = s_{12} \cong
\delta \sqrt{\sigma^2 + 1}/\epsilon \sigma \cong 0.07.
\end{array}
\end{equation}

\noindent
Note that the mixing $\theta^L_{23}$ is large and the mixing 
$\theta^L_{12}$ is small. As can be seen from Eq. (1), the smallness
of the angle $\theta^L_{12}$ is related to the smallness of the
corresponding angle for the quarks. The largeness of $\theta^L_{23}$,
on the other hand, is due to the large ``lopsided" entry $\sigma$.
There is no similar entry for the mixing of the first family. One 
could imagine introducing one, but one would find that doing so would
make it hard to fit many known quantities such as $V_{us}$, $V_{ub}$,
$m_e/m_{\mu}$, and $m_d/m_s$. It would seem, then, that this model
must give large $\nu_{\mu}-\nu_{\tau}$ mixing and small $\nu_e-
\nu_{\mu}$ mixing. However, as we shall now show, the same model 
actually can give bimaximal mixing.

We will suppose now that the up quark matrix element $U^0_{11}$ is not 
exactly zero, but is given by $\eta U^0_{33}$. If $m_u(1 \ {\rm GeV}) 
\approx 4 \ {\rm MeV}$, then $\eta \approx 6 \times 10^{-6}$,
which is a thousand times smaller than the smallest of the other model 
parameters, $\delta$ and $\delta'$. It is in this sense that the vanishing
of the first row and column of $U^0$ is an excellent approximation.
In $SO(10)$ the simplest possibility is that $N^0_{11}$ is also given
by $\eta M_U$. Thus, we assume for $N^0$ the form

\begin{equation}
N^0 = \left( \begin{array}{ccc}
\eta & 0 & 0 \\ 0 & 0 & - \epsilon \\
0 & \epsilon & 1 \end{array}
\right) M_U.
\end{equation}

\noindent
This tiny modification of Eq. (1) allows interesting effects for
certain forms of $M_R$.  What we find is that if the off-diagonal elements
in the first row and first column of $M_R$ are small or zero, then the 
small-angle MSW solution results as in \cite{ab2}, whereas if these 
elements are important, the large-angle solution of either the ``just-so''
vacuum or large-angle MSW oscillation type can result.  Instead of looking at
the most general forms for $M_R$, we will illustrate this in two 
representative cases:

\begin{equation}
\begin{array}{llll}
(I) & M_R^{I} = \left( \begin{array}{ccc}
B & 0 & 0 \\ 0 & 0 & A\\ 0 & A & 1 \end{array} \right) \Lambda_R, \;\; &
(II) & M_R^{II} = \left( \begin{array}{ccc}
0 & A & 0 \\ A & 0 & 0 \\ 0 & 0 & 1\end{array} \right) \Lambda_R. 
\end{array}
\end{equation}
 
\newpage

\noindent
{\bf Small-angle MSW Solution}

\noindent 
{\it Form (I):}  

\begin{equation}
M^{I}_{\nu} = - N^{0T} (M_R^{I})^{-1} N^0 = -\left( \begin{array}{ccc}
\eta^2/B & 0 & 0 \\ 0 & 0 & - \epsilon^2/A \\ 0 & - \epsilon^2/A & 
-2 \epsilon/A - \epsilon^2/A^2 \end{array} \right) \frac{M_U^2}{\Lambda_R}.
\end{equation}

\noindent
This light neutrino mass matrix is diagonalized by an orthogonal 
transformation of the type given in Eq. (3) for which Eq. (4) applies with 
$\theta^{\nu}_{23} \sim \epsilon$.  The presence of the new $\eta$ contribution 
to the 11 element of $N^0$ thus does not modify the small-angle MSW solution 
obtained earlier, if $M_R$ is of type (I).  The ratio of the 
mass differences required for this solution, $\delta m^2_{23} 
\cong 3.5 \times 10^{-3}\ {\rm eV^2}$, $\delta m^2_{12} \cong 7 \times 
10^{-6}\ {\rm eV^2}$, can easily be obtained with $A \simeq - \epsilon$, 
while the absolute mass scale follows with $M_U \sim 100$ GeV and $\Lambda_R
\sim 1.5 \times 10^{14}$ GeV. Moreover, the 
mixing results, $\sin^2 2\theta_{\mu\tau} \cong 1.0$ and 
$\sin^2 2\theta_{e\mu} \cong 0.008$, lie in the desired ranges for the 
atmospheric and small-angle MSW solutions \cite{atm,solar}.  

\vspace{0.5cm}

\noindent
{\bf Vacuum Oscillation Solution}

\noindent
{\it Form (II):}

\begin{equation}
M^{II}_{\nu} = - N^{0T} (M_R^{II})^{-1} N^0 = -\left( \begin{array}{ccc}
0 & 0 & - (\eta/A) \epsilon \\ 0 & \epsilon^2 & \epsilon \\
- (\eta/A) \epsilon & \epsilon & 1 \end{array} \right) \frac{M_U^2}{\Lambda_R}.
\end{equation}

\noindent
A rotation in the 2-3 plane by an angle $\tan \theta^{\nu}_{23} = 
\epsilon$ will eliminate the 22, 23, and 32 entries and induce
12 and 21 entries that are equal to $(\eta/A) \epsilon^2$ (neglecting
terms higher order in $\epsilon^2$). Following this with a rotation
in the 1-3 plane by an angle $\theta^{\nu}_{13} \cong (\eta/A) \epsilon$
brings the matrix to the form 

\begin{equation}
M'^{II}_{\nu} \cong - \left( \begin{array}{ccc}
- (\eta/A)^2 \epsilon^2 & (\eta/A) \epsilon^2 & 0 \\
(\eta/A) \epsilon^2 & 0 & 0 \\
0 & 0 & 1 \end{array} \right) \frac{M_U^2}{\Lambda_R}.
\end{equation}

\noindent
For $\left| \eta/A \right| \ll 1$, a pseudo-Dirac form for the mass matrix of 
$\nu_e$ and $\nu_{\mu}$ obtains with nearly-degenerate neutrinos.  One finds

\begin{equation}
\begin{array}{rl}
m_{\nu_e} \cong m_{\nu_{\mu}} &\cong \left| \eta/A
\right| \epsilon^2 (M_U^2/\Lambda_R),\\[0.1in]
\delta m^2_{12} &\cong 2 \left| \eta/A \right|^3 \epsilon^4
(M_U^2/\Lambda_R)^2,\\[0.1in]
\delta m^2_{23} &\cong m_{\nu_{\tau}}^2 \cong (M_U^2/\Lambda_R)^2\\
\end{array}
\end{equation} 

\noindent 
If one takes $\delta m^2_{12} \simeq 4 \times 10^{-10} eV^2$,
corresponding to the vacuum oscillation solution of the solar neutrino 
problem \cite{solar}, and $\delta m^2_{23} \simeq 3.5 \times 10^{-3} eV^2$ 
\cite{atm}, then
$\left| \eta/A \right| \simeq 0.05$ and the pseudo-Dirac condition is 
satisfied. This means that  $\left| \theta^{\nu}_{13} \right|
\simeq 7 \times 10^{-3}$, which we shall ignore, and 
$\left| \theta^{\nu}_{12} - \pi/4 \right| 
\cong 0.05$. Thus, to a good approximation we can write 
$\theta_{23}^{\nu} = \epsilon$, $\theta_{13}^{\nu} = 0$, and 
$\theta_{12}^{\nu} = \pi/4$, or

\begin{equation}
U_{\nu} = \left( \begin{array}{ccc}
1/\sqrt{2} & -1/\sqrt{2} & 0 \\
1/\sqrt{2} & 1/\sqrt{2} & \epsilon \\
-\epsilon/\sqrt{2} & -\epsilon/\sqrt{2} & 1 \end{array} \right). 
\end{equation}

\noindent
With the use of Eqs. (2) and (5), this gives for the lepton mixing matrix

\begin{equation}
V_{lepton} \cong 
\left( \begin{array}{ccc} 
1/\sqrt{2} & -1/\sqrt{2} & -s_{12} s'_{23} \\
c'_{23}/\sqrt{2} & c'_{23}/\sqrt{2} & -s'_{23} \\
s'_{23}/\sqrt{2} & - s'_{23}/\sqrt{2} & c'_{23} \end{array}
\right).
\end{equation}

\noindent 
with $s'_{23} \equiv \sin \theta_{\mu\tau}$, where
$\theta_{\mu \tau} \cong \theta^L_{23} - \epsilon
\cong 63^{\circ} - 8^{\circ} = 55^{\circ}$; hence $\sin^2 2\theta_{\mu\tau}
\simeq 0.9$ while $\sin^2 2\theta_{e\mu} \simeq 1.0$, which are consistent 
with the experimental limits.

It is interesting that the very small value of $\delta m^2_{12}$
needed for the vacuum oscillation solution to the solar neutrino problem
has a natural explanation in this approach. From Eq. (12)
it is apparent that $m_{\nu_{\mu}} \cong m_{\nu_e} \propto \eta$.
In other words, the very tiny parameter $\eta$ required to fit
$m_u/m_t$ --- a quantity pertaining to the {\it first} family ---
actually ends up controlling the masses of both the muon- and 
electron-neutrino which are nearly degenerate.  

\vspace{0.5cm}

\noindent
{\bf Large-angle MSW Solution}

\vspace{0.2cm}

The question arises whether form (II) can also give the
large-angle MSW solution to the solar neutrino problem \cite{solar}.
The answer is yes, though in this case $\sin^2 2 \theta_{e \mu}$
departs significantly from maximality. For this solution one needs to have 
$\Delta m^2_{12} \simeq 2 \times 10^{-5}\ {\rm eV^2}$ along with 
$\Delta m^2_{23} \simeq 3.5 \times 10^{-3}\ {\rm eV^2}$.  Although Eq. (11) 
still applies, the condition for pseudo-Dirac neutrinos is no longer 
satisfied as one finds $\eta/A \sim 1.8$.  Three Majorana neutrinos 
emerge for which 

\begin{equation}
\begin{array}{rl}
m_3 &\cong M^2_U/\Lambda_R,\\[0.1in]
m_2 &\cong {1\over{2}}(\eta/A)\epsilon^2 \left[\eta/A + \sqrt{4 + (\eta/A)^2}
	\right]M^2_U/\Lambda_R, \\[0.1in]
m_1 &\cong {1\over{2}}(\eta/A)\epsilon^2 \left[-\eta/A + \sqrt{4 + (\eta/A)^2}
	\right]M^2_U/\Lambda_R\\
\end{array}
\end{equation}

\noindent
With $M_U \sim 100$ GeV and $\Lambda_R \sim 1.7 \times 10^{14}$ GeV, the three
neutrino masses are given numerically  by $5.9 \times 10^{-2}\ {\rm eV},\ 
4.7 \times 10^{-3}$ eV and $9.3 \times 10^{-4}$ eV.
Making use of the above value for $\eta/A$, we can write down the equivalent of 
Eq. (13) for the neutrino mixing matrix 

\begin{equation}
U_{\nu} = \left( \begin{array}{ccc}
0.389 & -0.878 & -1.73\epsilon \\ 0.922 & 0.392 & 0.961\epsilon \\ 
-0.18\epsilon & - 2\epsilon & 0.961 
\end{array} \right).
\end{equation} 

\noindent
The lepton mixing matrix is then found numerically to be 

\begin{equation}
V_{lepton} \cong 
\left( \begin{array}{ccc} 
0.360 & -0.908 & -0.186 \\
0.469 & 0.367 & -0.812 \\
0.811 & 0.222 & 0.556 \end{array}
\right).
\end{equation}

\noindent
from which one obtains $\sin^2 2\theta_{\mu\tau} = 0.82$ and 
$\sin^2 2\theta_{e\mu} = 0.43.$  These values for the mixing angles
are on the low side of the allowed experimental region for this large-angle
MSW type of solution.  Moreover,
the $A \simeq 3 \times 10^{-6}$ parameter required is some thirty times 
smaller than that found earlier with a pair of pseudo-Dirac neutrinos; 
thus a considerably
larger hierarchy is required in the right-handed Majorana neutrino matrix
to reproduce the large-angle MSW mixing than for the vacuum solution.  
These features have also appeared in other forms we have assumed for $M_R$.

In summary, we have shown that, through the introduction of a small correction 
which gives mass to the up quark and hence also modifies the related Dirac 
neutrino matrix,
a bimaximal solution to the solar and atmospheric neutrino oscillations
can be achieved, provided the right-handed Majorana neutrino matrix mixes
the first family with the other two.  The large-angle vacuum solution 
is somewhat preferred over the large-angle MSW solution for the solar 
neutrino problem, since a smaller hierarchy is required in the Majorana
matrix and the mixing angles are more nearly maximal as suggested by 
present experimental data for those type of solutions.  But as presently 
understood, the model does not suggest a preference for the large-angle 
solutions over the small-angle MSW solution.

One of us (CHA) thanks the Fermilab Theoretical Physics Department for its
kind hospitality where much of his work was carried out.  The research of 
SMB was supported in part by the Department of Energy under contract No. 
DE-FG02-91ER-40626.  Fermilab is operated by Universities Research Association
Inc. under contract No. DE-AC02-76CH03000 with the Department of Energy.

\end{document}